\documentclass[12pt]{article}
\newcommand{\ba}{\begin{array}}
\newcommand{\ea}{\end{array}}

\newcommand{\C}{{\bf C}}

\newcommand{\del}{\partial}

\newcommand{\rar}{\rightarrow}

\newcommand{\scr}{\scriptsize}

\newcommand{\zb}{\bar{z}}

\makeatletter
\@addtoreset{equation}{section}

\makeatother

\begin{document}

\begin{titlepage}
\null
\begin{flushright}
%-/-
%\\
hep-th/0610006
\\
September, 2006
\end{flushright}

\vskip 1.5cm
\begin{center}

 {\Large \bf Notes on Exact Multi-Soliton Solutions of}

\vskip 0.5cm

{\Large \bf Noncommutative Integrable Hierarchies}

\vskip 1.7cm
\normalsize

{\large Masashi Hamanaka\footnote{The author visits Oxford 
from 16 August, 2005 to 15 December, 2006.\\
E-mail: hamanaka@math.nagoya-u.ac.jp}}

\vskip 1.5cm

        {\it Graduate School of Mathematics, Nagoya University,\\
                     Chikusa-ku, Nagoya, 464-8602, JAPAN}
\vskip 0.5cm

        {\it Mathematical Institute, University of Oxford,\\
                     24-29, St Giles', Oxford, OX1 3LB, UK}

\vskip 1.5cm

{\bf \large Abstract}

\vskip 0.5cm
 
\end{center}

We study exact multi-soliton solutions of
integrable hierarchies on noncommutative space-times
which are represented in terms of quasi-determinants
of Wronski matrices by Etingof, Gelfand and Retakh.
We analyze the asymptotic behavior of
the multi-soliton solutions and found that
the asymptotic configurations 
in soliton scattering process
can be all the same as commutative ones,
that is, the configuration of $N$-soliton solution has 
$N$ isolated localized energy densities
and the each solitary wave-packet preserves 
its shape and velocity in the scattering process.
The phase shifts are also the same as commutative ones.
Furthermore noncommutative toroidal
Gelfand-Dickey hierarchy is introduced
and the exact multi-soliton solutions are given.

\end{titlepage}
\clearpage
\baselineskip 6.15mm

\section{Introduction}

Extension of integrable systems and soliton theories
to non-commutative (NC) space-times
\footnote{In the present paper, the word ``NC'' always
refers to generalization to noncommutative spaces,
not to non-abelian and so on.}
have been studied by many authors for the last couple of years and 
various kind of integrable-like properties
have been revealed \cite{Integ, Lechtenfeld}.
This is partially motivated by recent developments
of NC gauge theories on D-branes. 
In the NC gauge theories, NC extension corresponds to
introduction of background magnetic fields and
NC solitons are, in some situations,
just lower-dimensional D-branes themselves.
Hence exact analysis of NC solitons just leads to
that of D-branes and various applications to D-brane dynamics
have been successful \cite{NC}. 
In this sense, NC solitons plays important roles
in NC gauge theories.

Most of NC integrable equations such as NC KdV equations
apparently belong not to gauge theories but to scalar theories.
However now, it is proved that they can be derived from NC 
anti-self-dual (ASD) Yang-Mills equations
by reduction \cite{Hamanaka06},
which is first conjectured explicitly
by the author and K.~Toda
\cite{HaTo}. (Original commutative one
is proposed by R.~Ward \cite{Ward}
and hence this conjecture is sometimes
called {\it NC Ward's conjecture}.)
Therefore analysis of exact soliton solutions of NC
integrable equations could be applied to
the corresponding physical situations
in the framework of N=2 string theory \cite{LPS,OoVa, Marcus}.

Furthermore, some soliton equations describe real phenomena
such as shallow water waves in fluid dynamics, optics and so on.
If noncommutativity in space-time affects soliton dynamics,
then we can check whether our universe is noncommutative or not
by comparing experimental results %in the asymptotic behavior 
and estimate the strength or the upper bound of the noncommutativity

Hence, construction and analysis of exact multi-soliton solutions
are worth studying from various viewpoints of integrable systems,
string theory, and perhaps detection of noncommutativity in our universe.

Exact multi-soliton solutions of noncommutative KP hierarchy are
constructed by Etingof, Gelfand and Retakh in 1997 \cite{EGR},
where quasi-determinants play crucial roles.
(For other applications of quasi-determinants
to noncommutative integrable systems, see e.g.
\cite{EGR2, GoVe, SaPe, Nimmo}.)
However, their discussion is general and explicit analysis of
the behavior of their soliton solutions has not yet been done.
Paniak also constructs multi-soliton solutions of
NC KP and KdV equations (not hierarchies)
and studies the scattering process \cite{Paniak}.\footnote{
Dimakis and M\"uller-Hoissen present perturbative corrections
with respect to a noncommutative parameter
in 2-soliton scatterings of the NC KdV equation \cite{DiMH_KdV}
before the Paniak's work.} 
However, the discussion about the soliton dynamics
is mainly focused on two-soliton scatterings.

In this paper, we study exact multi-soliton solutions of
NC integrable hierarchies in terms of quasi-determinants
of Wronski matrices, which is developed
by Etingof, Gelfand and Retakh.
We analyze the asymptotic behavior of
the multi-soliton solutions and found that
the asymptotic configurations can be real-valued
though NC fields take complex values in general. 
The behavior in soliton scatterings
is all the same as commutative ones,
            that is, the $N$-soliton solution
has $N$ isolated localized energy densities
and the each wave-packet preserve its shape and velocity
in the scattering process.
The phase shift is also the same as commutative one.

This paper is organized as follows.
In section 2, we make a brief introduction to
NC field theory in star-product formalism.
In section 3 and 4, we review definition and some properties of
quasi-determinants and their applications to construction
of multi-soliton solutions of NC integrable hierarchy
in star-product formalism. In the end of section 4,
we introduce NC toroidal
Gelfand-Dickey (GD) hierarchy and
give exact multi-soliton solutions which are new.
In section 5, we discuss the asymptotic behavior of them in detail.
Section 6 is devoted to 
conclusion and discussion.

\section{NC Field Theory in the star-product formalism}

NC spaces are defined
by the noncommutativity of the coordinates:
\begin{eqnarray}
\label{nc_coord}
[x^i,x^j]=i\theta^{ij},
\end{eqnarray}
where the constant $\theta^{ij}$ is 
called the {\it NC parameter}.
If the coordinates are real, NC parameters should be real.
%This relation looks like the canonical commutation
%relation in quantum mechanics
%and leads to ``space-space uncertainty relation.''
%Hence the singularity which exists on commutative spaces
%could resolve on NC spaces.
%This is one of the prominent features of NC
%field theories and yields various new physical objects.
Because the rank of the NC parameter is even,
dimension of NC space-times must be more than two.
Hence in this paper, we deal
not with integrable systems in $(0+1)$-dimension
such as the Painlev\'e equation,
but with ones in $(1+1)$ or $(2+1)$-dimension
such as the KdV and KP equations.
In $(1+1)$-dimension, we can take only space-time
noncommutativity as $[t,x]=i\theta$.
In $(2+1)$-dimension, there are essentially two kind of
choices of noncommutativity, that is,
space-space noncommutativity: $[x,y]=i\theta$
and space-time noncommutativity: $[t,x]=i\theta$
or $[t,y]=i\theta$, where the coordinates $(x,y)$ and $t$ 
correspond to space and time coordinates, respectively.

NC field theories are given by the replacement of ordinary products
in the commutative field theories with the {\it star-products} and
realized as deformed theories from the commutative ones.
The star-product is defined for ordinary fields on flat
spaces, explicitly by
\begin{eqnarray}
f\star g(x)&:=&\mbox{exp}
\left(\frac{i}{2}\theta^{ij} \partial^{(x^{\prime})}_i
\partial^{(x^{\prime\prime})}_j \right)
f(x^\prime)g(x^{\prime\prime})\Big{\vert}_{x^{\prime}
=x^{\prime\prime}=x}\nonumber\\
&=&f(x)g(x)+\frac{i}{2}\theta^{ij}\partial_i f(x)\partial_j g(x)
+{\cal{O}} (\theta^2),
\label{star}
\end{eqnarray}
where $\del_i^{(x^\prime)}:=\del/\del x^{\prime i}$
and so on. This explicit representation is known
as the {\it Moyal product} \cite{Moyal}.
The ordering of fields in nonlinear terms are determined
so that some structures such as gauge symmetries and Lax representations
should be preserved.

The star-product has associativity:
$f\star(g\star h)=(f\star g)\star h$,
and reduces to the ordinary product
in the commutative limit:  $\theta^{ij}\rar 0$.
The modification of the product  makes the ordinary
spatial coordinate ``noncommutative,''
that is, $[x^i,x^j]_\star:=x^i\star x^j-x^j\star x^i=i\theta^{ij}$.

We note that the fields themselves take c-number values
as usual and the differentiation and the integration for them
are well-defined as usual. 
A nontrivial point is that
NC field equations contain infinite number
of derivatives in general.
Hence the integrability of the equations
are not so trivial as commutative cases,
especially for space-time noncommutativity.

\section{Brief Review of Quasi-determinants}

In this section, we make a brief introduction
of quasi-determinants introduced by Gelfand and Retakh
\cite{GeRe, GeRe2} and present a few properties
of them which play important roles in the following sections.
The detailed discussion is seen in e.g. \cite{GGRW, KrLe}.
Relation between quasi-determinants and NC symmetric functions 
is seen in e.g. \cite{GKLLRT}.

Quasi-determinants are not just a generalization of
usual commutative determinants but rather
related to inverse matrices. From now on,
we suppose existence of all the inverses.

Let $A=(a_{ij})$ be a $N\times N$ matrix and 
$B=(b_{ij})$ be the inverse matrix of $A$,
that is, $A\star B=B\star A =1$.
Here all products of matrix elements are supposed to be
star-products, though the present discussion hold
for more general situation where the matrix elements
belong to a noncommutative ring.

Quasi-determinants of $A$ are defined formally
as the inverse of the elements of $B=A^{-1}$:
\begin{eqnarray}
 \vert A \vert_{ij}:=b_{ji}^{-1}.
\end{eqnarray}
In the commutative limit, this is reduced to
\begin{eqnarray}
 \vert A \vert_{ij} \longrightarrow
  (-1)^{i+j}\frac{\det A}{\det \tilde{A}^{ij}},
\label{limit}
\end{eqnarray}
where $\tilde{A}^{ij}$ is the matrix obtained from $A$
deleting the $i$-th row and the $j$-th column.

We can write down more explicit form of quasi-determinants.
In order to see it, let us recall the following formula
for a block-decomposed square matrix:
\begin{eqnarray*}
 \left(
 \begin{array}{cc}
  A&B \\C&D
 \end{array}
 \right)^{-1}
=\left(\begin{array}{cc}
% (A-BD^{-1}C)^{-1}&(C-DB^{-1}A)^{-1}\\(B-AC^{-1}D)^{-1}&(D-CA^{-1}B)^{-1}
 (A-B\star D^{-1}\star C)^{-1}
 &-A^{-1}\star B\star (D-C\star A^{-1} \star B )^{-1}\\
 -(D-C\star A^{-1}\star B)^{-1}\star C\star A^{-1}
&(D-C\star A^{-1}\star B)^{-1}
\end{array}\right),
\end{eqnarray*}
where $A$ and $D$ are square matrices.
We note that any matrix can be decomposed
as a $2\times 2$ matrix by block decomposition
where one of the diagonal parts is $1 \times 1$.
Then the above formula can be applied to the decomposed
$2\times 2$ matrix and an element of the inverse matrix is obtained.
Hence quasi-determinants can be also given iteratively by:
\begin{eqnarray}
 \vert A \vert_{ij}&=&a_{ij}-\sum_{i^\prime (\neq i), j^\prime (\neq j)}
  a_{ii^\prime} \star ((\tilde{A}^{ij})^{-1})_{i^\prime j^\prime} \star
  a_{j^\prime
  j}\nonumber\\
 &=&a_{ij}-\sum_{i^\prime (\neq i), j^\prime (\neq j)}
  a_{ii^\prime} \star (\vert \tilde{A}^{ij}\vert_{j^\prime i^\prime })^{-1}
  \star a_{j^\prime j}.
\end{eqnarray}

It is sometimes convenient to represent the quasi-determinant
as follows:
\begin{eqnarray}
 \vert A\vert_{ij}=
  \begin{array}{|ccccc|}
   a_{11}&\cdots &a_{1j} & \cdots& a_{1n}\\
   \vdots & & \vdots & & \vdots\\
   a_{i1}&~ & {\fbox{$a_{ij}$}}& ~& a_{in}\\
   \vdots & & \vdots & & \vdots\\
   a_{n1}& \cdots & a_{nj}&\cdots & a_{nn}
  \end{array}~.
\end{eqnarray}

Examples of quasi-determinants are,
for a $1\times 1$ matrix $A=a$
 \begin{eqnarray*}
  \vert A \vert  = a,
 \end{eqnarray*}
and 
for a $2\times 2$ matrix $A=(a_{ij})$
 \begin{eqnarray*}
  \vert A \vert_{11}=
   \begin{array}{|cc|}
   \fbox{$a_{11}$} &a_{12} \\a_{21}&a_{22}
   \end{array}
 =a_{11}-a_{12}\star a_{22}^{-1}\star a_{21},~~~
  \vert A \vert_{12}=
   \begin{array}{|cc|}
   a_{11} & \fbox{$a_{12}$} \\a_{21}&a_{22}
   \end{array}
 =a_{12}-a_{11}\star a_{21}^{-1}\star a_{22},\nonumber\\
  \vert A \vert_{21}=
   \begin{array}{|cc|}
   a_{11} &a_{12} \\ \fbox{$a_{21}$}&a_{22}
   \end{array}
 =a_{21}-a_{22}\star a_{12}^{-1}\star a_{11},~~~
  \vert A \vert_{22}=
   \begin{array}{|cc|}
   a_{11} & a_{12} \\a_{21}&\fbox{$a_{22}$}
   \end{array}
 =a_{22}-a_{21}\star a_{11}^{-1}\star a_{12}, 
 \end{eqnarray*}
 and for a $3\times 3$ matrix $A=(a_{ij})$
  \begin{eqnarray*}
  \vert A \vert_{11}
   &=&
   \begin{array}{|ccc|}
   \fbox{$a_{11}$} &a_{12} &a_{13}\\ a_{21}&a_{22}&a_{23}\\a_{31}&a_{32}&a_{33}
   \end{array}
=a_{11}-(a_{12}, a_{13})\star \left(
\begin{array}{cc}a_{22} & a_{23} \\a_{32}&a_{33}\end{array}\right)^{-1}
\star \left(
\begin{array}{c}a_{21} \\a_{31}\end{array}
\right)
\nonumber\\
%  &=&a_{11}-a_{12} ((\tilde{A}^{11})^{-1})_{22}  a_{21}
%           -a_{12} ((\tilde{A}^{11})^{-1})_{23}  a_{31}
%           -a_{13} ((\tilde{A}^{11})^{-1})_{32}  a_{21}
%           -a_{13} ((\tilde{A}^{11})^{-1})_{33}  a_{31}\nonumber\\
%  &=&a_{11}-a_{12} \vert \tilde{A}^{11} \vert_{22}  a_{21}
%           -a_{12} \vert \tilde{A}^{11} \vert_{32}  a_{31}
%           -a_{13} \vert \tilde{A}^{11} \vert_{23}  a_{21}
%           -a_{13} \vert \tilde{A}^{11} \vert_{33}  a_{31},\nonumber\\
  &=&a_{11}-a_{12}\star  \begin{array}{|cc|}
                   \fbox{$a_{22}$} & a_{23} \\a_{32}&a_{33}
                   \end{array}^{-1}  \star a_{21}
           -a_{12}\star \begin{array}{|cc|}
                   a_{22} & a_{23} \\\fbox{$a_{32}$}&a_{33}
                   \end{array}^{-1} \star a_{31}      \nonumber\\
&&~~~~    -a_{13}\star \begin{array}{|cc|}
                   a_{22} & a_{23} \\\fbox{$a_{32}$}&a_{33}
                                         \end{array}^{-1}\star  a_{21}
           -a_{13}\star \begin{array}{|cc|}
                   a_{22} & a_{23} \\a_{32}&\fbox{$a_{33}$}
                   \end{array}^{-1} \star a_{31},
 \end{eqnarray*}
and so on.

\vspace{3mm}

\section{Exact Soliton Solutions of NC Integrable Hierarchies}

In this section, we give exact multi-soliton
solutions of several NC integrable hierarchies
in terms of quasi-determinants.
In the commutative case, determinants of
Wronski matrices play crucial roles.
In the NC case, these determinants are just
replaced with the quasi-determinants.
We review foundation of the NC KP hierarchy and
the $l$-reduced hierarchies
(so called NC GD hierarchies or NC $l$KdV hierarchies),
and present the exact multi-soliton solutions of them
developed by Etingof, Gelfand and Retakh \cite{EGR}.
Finally we extend their discussion to the NC toroidal GD hierarchy.

\vspace{3mm}

An $N$-th order pseudo-differential operator $A$
is represented as follows
\begin{eqnarray}
 A=a_N \del_x^N + a_{N-1}\del_x^{N-1}+ \cdots
+ a_0 +a_{-1}\del_x^{-1}+a_{-2}\del_x^{-2}+\cdots,
\end{eqnarray}
where $a_i$ is a function of $x$ 
associated with noncommutative associative products
(here, the Moyal products).
When the coefficient of the highest order $a_N$ equals to 1,
we call it {\it monic}.
Here we introduce useful symbols:
\begin{eqnarray}
 A_{\geq r}&:=& \del_x^N + a_{N-1}\del_x^{N-1}+ \cdots + a_{r}\del_x^{r},\\
 A_{\leq r}&:=& A - A_{\geq r+1}
 = a_{r}\del_x^{r} + a_{r-1}\del_x^{r-1} +\cdots.%\\
% \res_{r} A &:=& a_{r}.
\end{eqnarray}
%The symbol $\res_{-1} A$ is especially called the {\it residue} of $A$.

The action of a differential operator $\partial_x^n$ on
a multiplicity operator $f$ is formally defined
as the following generalized Leibniz rule:
\begin{eqnarray}
 \partial_x^{n}\cdot f:=\sum_{i\geq 0}
\left(\begin{array}{c}n\\i\end{array}\right)
(\partial_x^i f)\partial^{n-i},
\end{eqnarray}
where the binomial coefficient is given by
\begin{eqnarray}
\label{binomial}
 \left(\begin{array}{c}n\\i\end{array}\right):=
\frac{n(n-1)\cdots (n-i+1)}{i(i-1)\cdots 1}.
\end{eqnarray}
We note that the definition of the binomial coefficient (\ref{binomial})
is applicable to the case for negative $n$,
which just define the action of
negative power of differential operators.
%The examples are,
%\begin{eqnarray}
% \partial_x^{-1}\cdot f&=&
%f\partial_x^{-1}-f^\prime\partial_x^{-2}
%+f^{\prime\prime}\partial_x^{-3}-\cdots,\nn
% \partial_x^{-2}\cdot f&=&
%f\partial_x^{-2}-2f^\prime\partial_x^{-3}
%+3f^{\prime\prime}\partial_x^{-4}-\cdots,\nn
% \partial_x^{-3}\cdot f&=&
%f\partial_x^{-3}-3f^\prime\partial_x^{-4}
%+6f^{\prime\prime}\partial_x^{-5}-\cdots,
%\end{eqnarray}
%where $\partial_x^{-1}$ in the RHS
%acts on a function as an integration $\int^x dx$.

The composition of pseudo-differential operators
is also well-defined and the total set
of pseudo-differential operators forms
an operator algebra.
For a monic pseudo-differential operator $A$, there exist
the unique inverse $A^{-1}$ and the unique $m$-th root $A^{1/m}$
which commute with $A$. (These proofs are all the same as commutative ones.)
For more on pseudo-differential operators
and Sato's theory, see e.g. \cite{Kupershmidt, Blaszak, Dickey, BBT}.

\vspace{3mm}

In order to define the NC KP hierarchy,
let us introduce a Lax operator:
\begin{eqnarray}
 L = \partial_x + u_2 \partial_x^{-1}
 + u_3 \partial_x^{-2} + u_4 \partial_x^{-3} + \cdots,
\end{eqnarray}
where the coefficients $u_k$ ($k=2,3,\ldots$) are functions
of infinite coordinates $\vec{x}:=(x_1,x_2,\ldots)$ with $x_1\equiv x$:
\begin{eqnarray}
 u_k=u_k(x_1,x_2,\ldots).
\end{eqnarray}
The noncommutativity is introduced into
the coordinates $(x_1,x_2,\ldots)$ as Eq. (\ref{nc_coord}) here.

The NC KP hierarchy is defined in Sato's framework as
\begin{eqnarray}
 \del_m L = \left[B_m, L\right]_\star,~~~m=1,2,\ldots,
\label{lax_sato}
\end{eqnarray}
where the action of $\del_m:=\del/\del x_m$
on the pseudo-differential operator $L$
should be interpreted to be coefficient-wise,
that is, $\del_m L :=[\del_m,L]_\star$ or $\del_m \del_x^k=0$.
The differential operator $B_m$ is given by
\begin{eqnarray}
 B_m:=(\underbrace{L\star \cdots \star L}_{ m{\scriptsize\mbox{
     times}}})_{\geq 0}=:(L^m)_{\geq 0}.
\end{eqnarray}
The KP hierarchy gives rise to a set of infinite differential
equations with respect to infinite kind of fields from the
coefficients in Eq. (\ref{lax_sato}) for a fixed $m$. Hence it
contains huge amount of differential (evolution) equations for all
$m$. The LHS of Eq. (\ref{lax_sato}) becomes $\del_m u_k$ which
shows a kind of flow in the $x_m$ direction. In the $x_2$-flow equations,
we can see that infinite kind of fields $u_3, u_4, u_5,\ldots$
are represented in terms of one kind of field  $u_2$ \cite{HaTo3, Hamanaka03}.

If we put the constraint $(L^l)_{\leq -1}=0$ or equivalently $L^l=B_l$ 
on the NC KP hierarchy (\ref{lax_sato}), 
we get a reduced NC KP
hierarchy which is called 
the {\it l-reduction} of the NC KP hierarchy, or
the {\it NC $l$KdV hierarchy}, or the $l$-th 
{\it NC Gelfand-Dickey hierarchy}. 
Especially, the 2-reduction of the NC KP hierarchy is just the NC KdV hierarchy.
Explicit examples are seen in e.g. \cite{Hamanaka03}.
(See also \cite{WaWa, OlSo, Wang}.)
%We can easily show
%\begin{eqnarray}
%\label{Nl}
%\frac{\partial u_k}{\partial x^{nl}}=0,
%\end{eqnarray}
%for all $n,k$ because
%\begin{eqnarray}
% \fr{dL^l}{dx^{nl}}=[B_{nl},L^l]=[(L^{l})^n,L^l]=0,
%\end{eqnarray}
%which implies Eq. (\ref{Nl}). 
%This time, the constraint $L^l=B_l$ gives simple relationships which make it
%possible to represent infinite kind of fields
%$u_{l+1},u_{l+2},u_{l+3},\ldots$ in terms of $(l-1)$ kind of
%fields $u_{2},u_{3},\ldots, u_{l}$. (cf. Appendix A in
%\cite{Hamanaka03}.) 

\vspace{3mm}

Now we construct multi-soliton solutions
of the NC KP hierarchy. 
Let us introduce the following functions,
\begin{eqnarray}
 \label{argument}
 f_s(\vec{x})=e_\star^{\xi(\vec{x};\alpha_s)}
  +a_s e_\star^{\xi(\vec{x};\beta_s)},
\end{eqnarray}
where
\begin{eqnarray}
 \xi(\vec{x};\alpha)=x_1\alpha+x_2 \alpha^2+x_3 \alpha^3+\cdots,
\end{eqnarray}
and $\alpha_s$, $\beta_s$ and $a_s$ are constants. 
Star exponential functions are defined by
\begin{eqnarray}
 e_\star^{f(x)}:=1+\sum_{n=1}^{\infty}\frac{1}{n!}
\underbrace{f(x)\star \cdots \star f(x)}_{n ~{\mbox{\scriptsize times}}}.
\end{eqnarray}

An $N$-soliton solution of the NC KP hierarchy (\ref{lax_sato})
is given by \cite{EGR},
\begin{eqnarray}
 L=\Phi_N \star \partial_x \Phi_N^{-1},
\label{Nsol1}
\end{eqnarray}
where
\begin{eqnarray}
 \Phi_N \star f&=&\vert W(f_1, \ldots, f_N, f)\vert_{N+1,N+1},\nonumber\\
&=&
\begin{array}{|ccccc|}
 f_1&f_2 & \cdots& f_N & f\\
 f^\prime_1&  f^\prime_2& \cdots& f^\prime_N&f^\prime\\
 \vdots& \vdots&\ddots & \vdots &\vdots\\
 f^{(N-1)}_1& f^{(N-1)}_2& \cdots & f^{(N-1)}_N &f^{(N-1)}\\
 f^{(N)}_1& f^{(N)}_2& \cdots &f^{(N)}_N & \fbox{$f^{(N)}$}\\
      \end{array}~ .
\label{sol_KP}
\end{eqnarray}
The Wronski matrix $W(f_1,f_2,\cdots, f_m)$ is given by
\begin{eqnarray}
 W(f_1,f_2,\cdots, f_m):=
\left(\begin{array}{cccc}
 f_1&f_2 & \cdots& f_m\\
 f^\prime_1& f^\prime_2& \cdots& f^\prime_m\\
 \vdots& \vdots&\ddots & \vdots\\
  f^{(m-1)}_1& f^{(m-1)}_2& \cdots & f^{(m-1)}_m\\
      \end{array}\right),
\end{eqnarray}
where $f_1,f_2, \cdots, f_m$ are functions of $x$
and $f^\prime:=\del f/\del x,~
f^{\prime\prime}:=\del^2 f/\del x^2,~
f^{(m)}:=\del^m f/\del x^m$ and so on. 

In the commutative limit, $\Phi_N \star f$ is reduced to
\begin{eqnarray}
 \Phi_N \star f \longrightarrow
  \frac{\det W(f_1,f_2,\ldots,f_N,f)}
{\det W(f_1,f_2,\ldots,f_N)},
\end{eqnarray}
which just coincides with commutative one \cite{Dickey}. 
In this respect, quasi-determinants are fit
to this framework of the Wronskian solutions.

{}From Eq. (\ref{Nsol1}), we have a more explicit form as 
\begin{eqnarray}
 u_2=\partial_x \left(\sum_{s=1}^{N} W^\prime_s \star W_s^{-1}  \right),
\label{Nsol2}
\end{eqnarray}
where
\begin{eqnarray}
 W_s:=\vert W(f_1,\ldots,f_s)\vert_{ss}.
\end{eqnarray}

The $l$-reduction condition $(L^l)_{\leq -1}=0$ or $L^l=B_l$
is realized at the level of the soliton solutions
by taking $\alpha_s^l=\beta_s^l$ or equivalently
$\alpha_s=\epsilon \beta_s$ for $s=1,\cdots,N$,
where $\epsilon$ is the $l$-th root of unity.

\vspace{3mm}

The present discussion is straightforwardly 
applicable for NC versions of
the matrix KP hierarchy \cite{KoOe, Blaszak, Dickey}, 
the toroidal (matrix) GD hierarchy
\cite{Bogoyavlenski,FTY,Billig, ISW, IkTa, IKT} and 
the (2-dimensional) Toda lattice hierarchy \cite{TaUe}
formulated by pseudo-differential operators,
because on commutative spaces, 
their exact soliton solutions are described by
determinants of (generalized) Wronski matrices.

For example, we can give exact $N$-soliton solutions
of the NC toroidal $l$KdV hierarchy
($l\geq 2$)\footnote{Toroidal $l$KdV hierarchy is
one of generalizations of $l$KdV hierarchy and 
first studied by Bogoyavlenskii \cite{Bogoyavlenski}
for $l=2$ and developed by Billig,
Iohara, Saito, Wakimoto, Ikeda and Takasaki
where the symmetry of the solution space is revealed to be
described in terms of a toroidal Lie algebra, that is, a central extension
of double loop algebra ${\cal G}^{\mbox{\scriptsize{tor}}}
={\cal{SL}}_l^{\mbox{\scriptsize{tor}}}$ \cite{Billig, ISW, IkTa}.
Hence we call it {\it toroidal lKdV hierarchy}
 or {\it toroidal GD hierarchy} in the present paper.}
which is defined as follows.

First, we introduce two kind of infinite variables
$\vec{x}=(x_1,x_2,\cdots)$ and $\vec{y}:=(y_0, y_l, y_{2l}, \cdots)$
with $(x,y)\equiv (x_1, y_0)$. Noncommutativity is introduced
into these coordinates.
Next let us define two kind of Lax operators with respect to $x$,
that is, an $l$-th order
differential operator $P=(L)^l_{\geq 0}$
and a 0-th order pseudo-differential operator $Q$,
where the coefficients depend on the two kind of
infinite variables. An differential operator $C_{ml}$
is also introduced in terms of $P$ and $Q$ as
$C_{ml}:=-(P^{m} \star Q)_{\geq 0}$. Then
we can obtain the NC toroidal $l$KdV hierarchy:
\begin{eqnarray}
 \frac{\del P}{\del x_{m}}=\left[B_m, P\right]_\star,&&~~~
 \frac{\del Q}{\del x_{m}}=\left[B_m, Q-\del_y\right]_\star,\\
 \frac{\del P}{\del y_{ml}}=\left[P^m\del_y+C_{ml}, P\right]_\star,&&~~~
 \frac{\del Q}{\del y_{ml}}=\left[P^m\del_y+C_{ml}, Q-\del_y\right]_\star.
\end{eqnarray}
For $l=2$, this includes the
NC Calogero-Bogoyavlenskii-Schiff equation \cite{Toda}.

The $N$-soliton solution is given by
\begin{eqnarray}
 P=\Phi_N \star \del_x^l \Phi_N^{-1},~~~Q=(\del_y\Phi_N) \star \Phi_N^{-1},
\end{eqnarray}
where %\phi:=\Phi_N \del_x^{-N}$ and
the arguments in $\Phi_N$ is modified as follows:
\begin{eqnarray}
 f_s(\vec{x},\vec{y})&:=&e_\star^{\xi_{r_s}(\vec{x},\vec{y};\alpha_s)}+a_s
  e_\star^{\xi_{r_s}(\vec{x},\vec{y};\beta_s)},\\
 \xi_r(\vec{x},\vec{y};\alpha)
  &:=&\xi(\vec{x};\alpha)+r\xi(\vec{y};\alpha)\nonumber\\
  &=&x_1\alpha+x_2 \alpha^2+x_3 \alpha^3+\cdots+ry_0+ry_{l}
   \alpha^l+ry_{2l} \alpha^{2l}+\cdots,
\end{eqnarray}
with $\alpha_s^l=\beta_s^l$, where $r$ is a constant.
The proof is the same as
the commutative one. (For the details,
see section 5.1 in \cite{IkTa}.)
A key point of the proof is to
show the evolution equations of $\Phi_N$:
\begin{eqnarray*}
 \frac{\partial \Phi_N}{\partial x_m}
  &=& - (\Phi_N\partial_x^{m}\Phi_N^{-1})_{\leq -1}
  \star \Phi_N =  B_m \star \Phi_N -\Phi_N \partial_x^{m},\\
 \frac{\partial \Phi_N}{\partial y_{ml}}
  &=& (P^m \star Q)_{\leq -1} \star \Phi_N
  =(P^m\partial_y+C_{ml})\star \Phi_N,
\end{eqnarray*}
where the following property of quasideterminant
plays crucial roles:
\begin{eqnarray*}
\Phi_N \star f_s=\vert
W(f_1, \ldots, f_N, f_s)\vert_{N+1,N+1}=0,~~~{\mbox{for}}~~~ s=1,\cdots,N. 
\end{eqnarray*}
This hierarchy generally gives rise to
$(2+1)$-dimensional integrable equations
where space and time coordinates are
$(x,y)$ and some other coordinate, respectively.

\section{Asymptotic Behavior of the Exact Soliton Solutions}

In this section, we discuss asymptotic behavior of
the multi-soliton solutions at spatial infinity
or infinitely past and future. In the star-product formalism,
all coordinates are regarded as c-numbers and
hence we can plot the configurations as usual and
interpret the positions of localized wave packets,
effect of coordinate shifts and so on as usual.
Here we restrict ourselves to NC KdV and KP hierarchies,
however, this observation would be also applicable
to other NC hierarchies.

First, we present some special properties of the star
exponential functions relevant to
behavior of NC soliton solutions. In this section,
we restrict ourselves to a specific equation
on $(2+1)$ or $(1+1)$-dimensional space-time
and noncommutativity should be introduce to
some two specific space-time coordinates.
Let us suppose that the specified NC coordinates
are denoted by $x_i$ and $x_j$ ($i<j$)
which satisfies $[x_i,x_j]_\star=i\theta$.

First, let us comment on an important formula 
which is relevant to one-soliton solutions.
Defining new coordinates $z:=x_i+v x_j,
\zb:=x_i-v x_j$, we can easily see
\begin{eqnarray}
 f(z)\star g(z)= f(z) g(z)
\end{eqnarray}
because the Moyal-product (\ref{star}) is rewritten in terms of
$(z,\zb)$ as \cite{DiMH_NLS}
\begin{eqnarray}
 f(z,\zb)\star g(z,\zb)=
e^{iv\theta\left(
\partial_{\zb^\prime}
\partial_{z^{\prime\prime}}-
\partial_{z^\prime}
\partial_{\zb^{\prime\prime}}
\right)}f(z^\prime,\zb^\prime)
g(z^{\prime\prime},\zb^{\prime\prime}) \Big{\vert}_{\scr
\begin{array}{c} z^{\prime}
=z^{\prime\prime}=z\\
\zb^{\prime} =\zb^{\prime\prime}=\zb. \end{array}}
\label{hol}
\end{eqnarray}
Hence NC one soliton-solutions are essentially the same as
commutative ones for both space-time
and space-space noncommutativity cases.

\vspace{3mm}

When $f(x)$ is a linear function,
the treatment of $e_\star^{f(x)}$ becomes easy as follows:
\begin{eqnarray}
\label{inverse_exp}
 (e_\star^{\xi(\vec{x};\alpha)})^{-1} &=& e_\star^{-\xi(\vec{x};\alpha)},\\
 \partial_x e_\star^{\xi(\vec{x};\alpha)}
  &=& \alpha e_\star^{\xi(\vec{x};\alpha)}.
\label{der_exp}
\end{eqnarray}
The proof can be seen as well from the fact that
because of Eq. (\ref{hol}), the star exponential function
of a linear function itself reduces to commutative one, that is,
$e_\star^{\xi(\vec{x},\alpha)}=e^{\xi(\vec{x},\alpha)}$.
These formula are crucial in discussion on
asymptotic behavior of $N$-soliton solutions.

Furthermore, the Baker-Campbell-Hausdorff (BCH) formula implies
\begin{eqnarray}
 e_\star^{\xi(\vec{x};\alpha)}\star e_\star^{\xi(\vec{x};\beta)}
=e^{(i/2)\theta(\alpha^i\beta^j-\alpha^j\beta^i)}
e_\star^{\xi(\vec{x};\alpha)+\xi(\vec{x};\beta)}.
=e^{i\theta (\alpha^i \beta^j-\alpha^j\beta^i)}
e_\star^{\xi(\vec{x};\beta)}\star e_\star^{\xi(\vec{x};\alpha)}.
\label{BCH}
\end{eqnarray}
The factor 
$(i/2)\theta (\alpha^i \beta^j-\alpha^j\beta^i)$ can
be absorbed by a coordinate shift in $\xi(\vec{x};\alpha)$,
and hence there is a possibility that noncommutativity
might affect coordinate shifts by the factor such as phase shifts
in the asymptotic behavior.
When coordinates and fields are treated as complex,
such a coordinate shift by a complex number causes no problem.
However, if we want to apply NC integrable equations
to real phenomena, such as, shallow water waves,
then it becomes hard to interpret physically.
Let us see what happens in the asymptotic region.

\subsection{Asymptotic behavior of NC KdV hierarchy}

First, let us discuss the NC KdV hierarchy and 
the asymptotic behavior of the $N$-soliton solutions.
The NC KdV hierarchy is the 2-reduction of the NC KP hierarchy
and realized by putting $\beta_s=-\alpha_s$ on the 
$N$-soliton solutions of the NC KP hierarchy.
Here the constants $\alpha_s$ and $a_s$ are non-zero real numbers
and $a_s$ is positive.
Because of the permutation property of the columns of quasi-determinants
(cf. section 1.1 in \cite{GeRe}),
we can assume $\alpha_1< \alpha_2 <\cdots <\alpha_N$.

In the NC KdV hierarchy, the $x^{2n}$-th flow becomes trivial
and in the $x^{2n+1}$-th flow equation, space and time
coordinates are specified as $(x,t)\equiv(x_1,x_{2n+1})$.
%Here we restrict ourselves on the $n=1$ case 
%where the equation is just the NC KdV equation, because
%the discussion can be straightforwardly extended to other $n$.

Now let us define a new coordinate $\tilde{x}:=x+\alpha_I^{2n} t$
comoving with the $I$-th soliton and take $t\rightarrow \pm \infty$
limit.\footnote{Such kind of observation for soliton scatterings
in NC integrable equations is first seen in \cite{LePo}.
 (See also \cite{Lechtenfeld}.)}
Then, because of $x+\alpha_s^{2n} t=x+\alpha_I^{2n} t + (\alpha_s^{2n}
-\alpha_I^{2n})t$,
either $e_\star^{\alpha_{s}(x + \alpha_{s}^{2n} t)}$
or $e_\star^{-\alpha_{s}(x + \alpha_{s}^{2n} t)}$
goes to zero for $s\neq I$.
Hence the behavior of $f_s$ becomes at $t\rightarrow +\infty$:
\begin{eqnarray}
 f_s(\vec{x})\longrightarrow
\left\{\begin{array}{ll}
a_s e_\star^{- \alpha_{s}(x + \alpha_{s}^{2n} t)} &s< I\\
e_\star^{\alpha_{I}(x + \alpha_{I}^{2n} t)}
+a_I e_\star^{-\alpha_{I}(x + \alpha_{I}^{2n} t)}&s=I\\
 e_\star^{ \alpha_{s}(x + \alpha_{s}^{2n} t)} &s> I,\\
       \end{array}
\right. 
\end{eqnarray}
and at $t\rightarrow -\infty$:
\begin{eqnarray}
 f_s(\vec{x})\longrightarrow
\left\{\begin{array}{ll}
 e_\star^{ \alpha_{s}(x + \alpha_{s}^{2n} t)} &s< I\\
e_\star^{\alpha_{I}(x + \alpha_{I}^{2n} t)}
+a_I e_\star^{-\alpha_{I}(x + \alpha_{I}^{2n} t)}&s=I\\
a_s e_\star^{- \alpha_{s}(x + \alpha_{s}^{2n} t)} &s> I.\\
       \end{array}
\right. 
\end{eqnarray}
We note that the $s$-th ($s\neq I$) column
is proportional to a single 
exponential function
$e_\star^{\pm \alpha_{s}(x + \alpha_{s}^{2} t)}$ due to
Eq. (\ref{der_exp}).
Because of the multiplication property of
columns of quasi-determinants (cf. section 1.2 in \cite{GeRe}),
we can eliminate a common invertible
factor from the $s$-th column in $\vert A\vert_{ij}$ where $s\neq j$.
(Note that this exponential function is actually invertible
as is shown in Eq. (\ref{inverse_exp}).)
Hence the $N$-soliton solution becomes the
following simple form where only the $I$-th column
is non-trivial, at $t\rightarrow +\infty$:
\begin{eqnarray*}
 \Phi_N \star f\rightarrow~~~~~~~~~~~~~~~~~~~~~~~~~~~~~~~~~~~~~~~~~~~~~~~~~~~~~~~~~~~~~~~~~~~~~~~~~~~~~~~~~~~~~~~~~~~~~~~~~~~~~~~~~~~~~  \nonumber\\
\begin{array}{|cccccccc|}
1 & \cdots&1& e_\star^{\xi(\vec{x};\alpha_I)}
+a_I e_\star^{-\xi(\vec{x};\alpha_I)} &1&\cdots&1 & f\\
 -\alpha_1& \cdots&-\alpha_{I-1}
  &\alpha_I(e_\star^{\xi(\vec{x};\alpha_I)}
-a_I e_\star^{-\xi(\vec{x};\alpha_I)})&\alpha_{I+1}& \cdots&
\alpha_N  &f^\prime\\
 \vdots&  &\vdots&\vdots &\vdots& & \vdots &\vdots\\
\! (-\alpha_1)^{N-1}\!&\! \cdots\!&(-\alpha_{I-1})^{N-1}\!&\! 
\alpha_I^{N-1}(e_\star^{\xi(\vec{x};\alpha_I)}
+(-1)^{N-1}
a_I e_\star^{-\xi(\vec{x};\alpha_I)})\!
&\alpha^{N-1}_{I+1}\!&\!\cdots\! &\! \alpha_N^{N-1} \!&\! f^{(N-1)}\\
 (-\alpha_1)^{N}& \cdots&(\alpha_{I-1})^{N}& 
\alpha_I^N(e_\star^{\xi(\vec{x};\alpha_I)}
+(-1)^N a_I e_\star^{-\xi(\vec{x};\alpha_I)})
 &\alpha^{N}_{I+1}&\cdots&\alpha_N^{N}& \fbox{$f^{(N)}$}\\
\end{array}~ ,
\end{eqnarray*}
and at $t\rightarrow -\infty$:
\begin{eqnarray*}
 \Phi_N \star f\rightarrow~~~~~~~~~~~~~~~~~~~~~~~~~~~~~~~~~~~~~~~~~~~~~~~~~~~~~~~~~~~~~~~~~~~~~~~~~~~~~~~~~~~~~~~~~~~~~~~~~~~~~~~~~~~~~ \\
\begin{array}{|cccccccc|}
1 & \cdots&1& e_\star^{\xi(\vec{x};\alpha_I)}
+a_I e_\star^{-\xi(\vec{x};\alpha_I)} &1&\cdots&1 & f\\
 \alpha_1& \cdots&\alpha_{I-1}
  &\alpha_I(e_\star^{\xi(\vec{x};\alpha_I)}
-a_I e_\star^{-\xi(\vec{x};\alpha_I)})&-\alpha_{I+1}& \cdots&
-\alpha_N  &f^\prime\\
 \vdots&  &\vdots&\vdots &\vdots& & \vdots &\vdots\\
 \alpha_1^{N-1}\!&\! \cdots\!&\!\alpha^{N-1}_{I-1}\!&\! 
\alpha_I^{N-1}(e_\star^{\xi(\vec{x};\alpha_I)}
+(-1)^{N-1}
a_I e_\star^{-\xi(\vec{x};\alpha_I)})\!
&(-\alpha_{I+1})^{N-1}\!&\!\cdots\! &\! (-\alpha_N)^{N-1}\! &\!f^{(N-1)}\\
 \alpha_1^{N}& \cdots&\alpha^{N}_{I-1}& 
\alpha_I^N(e_\star^{\xi(\vec{x};\alpha_I)}
+(-1)^N a_I e_\star^{-\xi(\vec{x};\alpha_I)})
 &(-\alpha_{I+1})^{N}&\cdots&(-\alpha_N)^{N}& \fbox{$f^{(N)}$}\\
\end{array}~ .
\end{eqnarray*}
Here we can see that all elements in between the first column and the
$N$-th column commute and depend only on $x + \alpha_{I}^{2n} t$
in $\xi(\vec{x};\alpha_I)$, 
which implies that the corresponding asymptotic
configuration coincides with the commutative one,\footnote{
Note that because $f$ is arbitrary,
there is no need to consider the products between
a column and the $(N+1)$-th column.
This observation for asymptotic behavior
can be made from Eq. (\ref{Nsol2}) also.}
that is, the $I$-th one-soliton configuration with some
coordinate shift so called the {\it phase shift}.
The commutative discussion has been studied in this way 
by many authors, and therefore, we conclude that
for the NC KdV hierarchy,
{\it asymptotic behavior of the multi-soliton solutions 
is all the same as commutative one}, and as the results,
{\it the $N$-soliton solutions possess $N$ localized energy densities
and in the scattering process, they never decay and
preserve their shapes and velocities of the localized solitary waves.
The phase shifts also occur by the same degree as commutative ones}.

\vspace{3mm}

Finally, we make a brief comment on the 2-soliton solutions.
In this situation, space-time dependence appears only as
two kind of exponential factors $e^{\pm \alpha_1 (x+\alpha_1^{2n} t)}$
and $e^{\pm \alpha_2 (x+\alpha_2^{2n} t)}$.
Noncommutativity of them could have effects by the factor
$e^{\pm (i/2)\alpha_1\alpha_2(\alpha_1^{2n}-\alpha_2^{2n})\theta}$
because of the BCH formula.
However, if the two kind of parameters satisfy
\begin{eqnarray}
 \label{2sol}
 \frac{1}{2}\alpha_1\alpha_2(\alpha_1^{2n}-\alpha_2^{2n})\theta =2\pi k,
\end{eqnarray}
where $k$ is an non-zero integer, then,
the effects of noncommutativity perfectly disappear
at the every stage of calculations
and the behavior of the 2-soliton
solution perfectly coincides with that of commutative one
at any time and any location. However the condition
(\ref{2sol}) is given specially by hand, and
the mathematical and physical meaning
of this observation is still unknown.
%At least, for $n=1$, there exists appropriate values of
%$\alpha_1$ and $\alpha_2$ which satisfy Eq. (\ref{2sol}),
%as is plotted in Fig 1.

\subsection{Asymptotic behavior of NC KP hierarchy}

Now, let us discuss the NC KP hierarchy and 
the asymptotic behavior of the $N$-soliton solutions.
%We can assume $\alpha_1< \alpha_2 <\cdots <\alpha_N$
%without loss of generality.
The space and time
coordinates are $(x,y,t)\equiv(x_1,x_2,x_{n})$
and noncommutativity is introduced into
some specified two coordinates among $x,y$ and $t$.
The specified NC coordinates are also denoted by $x_i$
and $x_j$ with $[x_i,x_j]_\star=i\theta$.
Here the constants $\alpha_s$ and $\beta_s$
are non-zero real numbers and the constant $a_s$
will be redefined later. 

As we mentioned at the beginning of the present section,
one-soliton solutions are all the same as commutative ones.
However, we have to treat carefully for the NC KP hierarchy.
{}From Eq. (\ref{Nsol2}), naive one-soliton solution
can be expressed as follows 
\begin{eqnarray}
 u_2&=&\partial_x\left(\partial_x(e_\star^{\xi(\vec{x};\alpha)}
+a e_\star^{\xi(\vec{x};\beta)})\star (e_\star^{\xi(\vec{x};\alpha)}
+a e_\star^{\xi(\vec{x};\beta)})^{-1}\right)\nonumber\\
 &=&
  \partial_x\left((\alpha^i+a\beta^i
\Delta e_\star^{\eta(\vec{x};\alpha,\beta)})\star
(1+a \Delta
e_\star^{\eta(\vec{x};\alpha,\beta)})^{-1}\right),
\end{eqnarray}
where
\begin{eqnarray}
 \eta(\vec{x};\alpha,\beta)&:=&x(\beta-\alpha)+y(\beta^2-\alpha^2)
  +t(\beta^n-\alpha^n)\nonumber\\
\Delta&:=&e^{\frac{i}{2}\theta(\alpha^i\beta^j-\alpha^j\beta^i)}.
\end{eqnarray}
We note that the factor $\Delta$
can be absorbed by redefining a coordinate
such as $x\rightarrow x+(\beta-\alpha)^{-1}
(i/2)\theta(\alpha^i\beta^j-\alpha^j\beta^i)$.
The final form of the solution depend only on
$x_i(\beta_I^i-\alpha_{I}^{i})+x_j(\beta_I^j-\alpha_{I}^{j})$
for NC coordinates and the Moyal products disappear.
Hence there becomes no dependence of complex numbers,
and the one-soliton solution is the same as commutative one
in this sense. However now we treat the coordinates as real
and it would be better to redefine a positive real number
$\tilde{a}$ which satisfies $a=\tilde{a}\Delta^{-1}$,
so that $f_1=e_\star^{\xi(\vec{x};\alpha)}
+a e_\star^{\xi(\vec{x};\beta)}=\left(
1+\tilde{a} e_\star^{\eta(\vec{x};\alpha,\beta)}\right)
\star e_\star^{\xi(\vec{x};\alpha)}$,
in order to avoid such a coordinate shift by a complex number.

This point becomes important for the multi-soliton solutions.
The constants $a_s$ in the $N$-soliton solution of the NC KP hierarchy
should be replaced with a positive real number
$\tilde{a}_s$ which satisfies $a_s=\tilde{a}_s\Delta_s^{-1}$
where $\Delta_s:=e^{(i/2)\theta( \alpha_s^i \beta_s^j-\alpha_s^j\beta_s^i)}$,
because the $N$-soliton configuration reduces to
a ($I$-th) one-soliton configuration
when we set $\alpha_s=\beta_s=0$ for all $s (\neq I)$.

Let us define new coordinates
comoving with the $I$-th soliton as follows:
\begin{eqnarray}
p:=x+\alpha_I y+\alpha_I^{n-1} t,~~~
q:=x+\beta_I y+\beta_I^{n-1} t.
\label{comoving}
\end{eqnarray}
Then the function $\xi(x,y,t;\alpha_s)$ can be rewritten
in terms of the new coordinates
as $\xi(p,q,x_r;\alpha_s)=A(\alpha_s)p+B(\alpha_s)q+C(\alpha_s)x_r$
where $x_r$ is a specified coordinate among $x,y$ and $t$,
and $A(\alpha_s), B(\alpha_s)$ and $C(\alpha_s)$ are
real constants depending on $\alpha_I,\beta_I$ and $\alpha_s$.
For example, in the case of $x_r\equiv t$, we can get from Eq.
(\ref{comoving})
\begin{eqnarray}
 \left(
 \begin{array}{c}x\\y\end{array}
 \right)=\frac{1}{\beta_I-\alpha_I}
  \left(
 \begin{array}{c}
 \beta_Ip-\alpha_Iq+\alpha_I\beta_I(\beta_I^{n-2}-\alpha_I^{n-2})t\\
 -p+q+(\alpha_I^{n-1}-\beta_I^{n-1})t
 \end{array}
 \right),
\end{eqnarray}
and find
\begin{eqnarray*}
 \xi\!&=&\! x+\alpha_sy+\alpha_s^{n-1}t\\
  \! &=&\! \frac{\beta_I-\alpha_s}{\beta_I-\alpha_I}p
  +\frac{\alpha_s-\alpha_I}{\beta_I-\alpha_I}q
  +\left\{\alpha_s^{n-1}+\frac{
   \alpha_I\beta_I(\beta_I^{n-2}-\alpha_I^{n-2})
  +\alpha_s(\alpha_I^{n-1}-\beta_I^{n-1})}{\beta_I-\alpha_I}
 \right\}t.
\end{eqnarray*}
Here we suppose that $C(\alpha_s)\neq C(\beta_s)$
which corresponds to pure soliton scatterings.\footnote{
The condition $C(\alpha_s)= C(\beta_s)$ could
lead to soliton resonances in commutative case.}

Now let us take $x_{r}\rightarrow \pm \infty$ limit,
then, for the same reason as in the NC KdV hierarchy,
we can see that the asymptotic behavior of $f_s$ becomes:
\begin{eqnarray}
 f_s(\vec{x})\longrightarrow
\left\{\begin{array}{ll}
A_s e_\star^{\xi(\vec{x};\gamma_{s})} &s\neq I\\
e_\star^{\xi(\vec{x};\alpha_{I})}
+a_I e_\star^{\xi(\vec{x};\beta_{I})}&s=I
\end{array}
\right.
\end{eqnarray}
where $A_s$ is some real constant whose value is 1 or $a_s$,
and $\gamma_s$ is a real constant
taking a value of $\alpha_s$ or $\beta_s$.
As in the case of the NC KdV hierarchy,
the $s$-th ($s\neq I$) column
is proportional to a single exponential function
and we can eliminate this factor from the $s$-th column. 
Hence in the asymptotic region $x_r\rightarrow \pm \infty$,
the $N$-soliton solution becomes the
following simple form where only the $I$-th column
is non-trivial:
\begin{eqnarray*}
 \Phi_N \star f &\rightarrow& 
\begin{array}{|cccccccc|}
 1& \cdots &1& e_\star^{\xi(\vec{x};\alpha_{I}) }
+a_I e_\star^{\xi(\vec{x};\beta_{I}) } &1&\cdots&1 & f\\
 \gamma_1& \cdots &\gamma_{I-1}&\alpha_Ie_\star^{\xi(\vec{x};\alpha_{I}) }
+a_I\beta_I e_\star^{\xi(\vec{x};\beta_{I}) } &\gamma_{I+1}& \cdots&
\gamma_N  &f^\prime\\
 \vdots& & \vdots&\vdots &\vdots& & \vdots &\vdots\\
 \gamma_1^{N-1}& \cdots &\gamma^{N-1}_{I-1}& 
\alpha_I^{N-1}e_\star^{\xi(\vec{x};\alpha_{I}) }
+
a_I \beta_I^{N-1}
e_\star^{\xi(\vec{x};\beta_{I})}  &\gamma^{N-1}_{I+1}
&\cdots & \gamma_N^{N-1} &f^{(N-1)}\\
 \gamma_1^{N}& \cdots &\gamma^{N}_{I-1}& 
\alpha_I^Ne_\star^{\xi(\vec{x};\alpha_I) }
+a_I \beta_I^N e_\star^{\xi(\vec{x};\beta_{I}) }
 &\gamma^{N}_{I+1} &\cdots& \gamma_N^{N}& \fbox{$f^{(N)}$}\\
\end{array}~ \nonumber\\
&&=\begin{array}{|cccccccc|}
1& \cdots &1& 1
+\tilde{a}_I e_\star^{\eta(\vec{x};\alpha_I,\beta_I) }&1 &\cdots&1& f\\
  \gamma_1 & \cdots& \gamma_{I-1}&\alpha_I
+\tilde{a}_I\beta_I e_\star^{\eta(\vec{x};\alpha_I,\beta_{I}) } &\gamma_{I+1}
& \cdots&
\gamma_N  &f^\prime\\
 \vdots& & \vdots&\vdots &\vdots & & \vdots &\vdots\\
 \gamma_1^{N-1}& \cdots &\gamma^{N-1}_{I-1}& 
\alpha_I^{N-1}
+
\tilde{a}_I \beta_I^{N-1}
e_\star^{\eta(\vec{x};\alpha_I,\beta_{I})}
 &\gamma^{N+1}_{I+1}&\cdots & \gamma_N^{N-1} &f^{(N-1)}\\
  \gamma_1^{N}& \cdots &\gamma^{N}_{I-1}& 
\alpha_I^N
+\tilde{a}_I \beta_I^N e_\star^{\eta(\vec{x}; \alpha_I,\beta_{I}) }
  &\gamma^{N}_{I+1}&\cdots& \gamma_N^{N}& \fbox{$f^{(N)}$}\\
\end{array}~.
\end{eqnarray*}
Here we can see that all elements between the first column and the
$N$-th column commute and depend only on
$x_i(\beta_I^i-\alpha_{I}^{i})+x_j(\beta_I^j-\alpha_{I}^{j})$
for NC coordinates, 
which implies that the corresponding asymptotic
configuration coincides with the commutative one.
Hence, we can also conclude that for the NC KP hierarchy,
{\it asymptotic behavior of the multi-soliton solutions 
is all the same as commutative one in the process of
pure soliton scatterings},
and as the results,
{\it the $N$-soliton solutions possess $N$ localized energy densities
and in the pure scattering process, they never decay and
preserve their shapes and velocities of localized solitary waves.}
Asymptotic behavior of two-soliton solution of NC KP equation
studied by Paniak \cite{Paniak} actually coincides with
our result for $n=3, N=2$.

\vspace{3mm}

Now we restricted ourselves to the NC KP hierarchy, however,
this observation would be also true of other kind of NC hierarchies,
such as, the (2-dimensional) NC Toda lattice hierarchy \cite{Sakakibara},
the NC toroidal GD hierarchy presented in section 4
and the NC matrix KP hierarchy \cite{DiMH_hie}
because the soliton solutions could be represented by
such kind of (generalized) Wronski matrices here,
and the asymptotic analysis would be almost the same.

\section{Conclusion and Discussion}

In this paper, we studied exact multi-soliton
solutions of NC integrable hierarchies, including
NC KP and toroidal KP hierarchies and the reductions,
in terms of quasi-determinants.
We found that the asymptotic behavior of them
could be all the same as commutative ones
in the process of (pure) soliton scatterings.
This implies that the exact soliton solutions are actually 
solitons in the sense that the configuration
has localized energy densities and never decay,
and the phase shifts also appear
by the same degree as in the commutative case.

It would be reasonable that there is no difference
in asymptotic behavior of pure soliton scatterings
on between commutative and NC spaces,
because in asymptotic region,
star-products reduce to ordinary commutative products
and the effect of noncommutativity disappears.
These results imply that
we cannot detect effects of noncommutativity of space-time
by observing such soliton dynamics.
However, total behavior of them is unknown
and it is worth studying further to find
different aspects of the NC soliton dynamics from commutative ones.

Dynamics in soliton resonances is also interesting.
{}From the present results of pure soliton scatterings,
we could naturally expect that the configurations in soliton resonances
would not be affected by noncommutativity in asymptotic region
though we might need to make further modifications
in the multi-soliton solutions
such as $a_s=\tilde{a}_s \Delta_s^{-1}$ for the NC KP equation.
Quantum treatments of the soliton scatterings is also
interesting, such as properties of factorized S-matrix
of NC sine-Gordon model \cite{LMPPT}.
Furthermore, the existence of multi-soliton solutions is
important in integrable systems and the present observations
might be a hint to reveal NC Hirota's bilinearization,
theory of tau-functions and the structure of solution spaces.

\subsection*{Acknowledgments}

The author would like to thank
C.~S.~Chu, H. Kajiura, L.~Mason, S.~Moriyama,
Y.~Ohta, V.~Retakh, T.~Suzuki, K.~Takasaki, K.~Toda
and J.~P.~Wang for useful comments.
He is also grateful to 
C.~R.~Gilson, J.~J.~C.~Nimmo and I.~Strachan
for warm hospitality and fruitful discussion
during stay 
at the department of mathematics, university of Glasgow.
This work was partially supported
by the Daiko foundation (\#9095),
and the Yamada Science Foundation
for the promotion of the natural science,
and Grant-in-Aid for Young Scientists (\#18740142).

\end{document}